\documentclass[pra,twocolumn,showpacs,amsmath,amssymb,superscriptaddress,longbibliography]{revtex4-1}
\usepackage{psfrag,graphicx}
\usepackage{amsfonts,amssymb,amsmath}      
\usepackage{graphicx}
\usepackage{dcolumn}
\usepackage{bm}
\usepackage{float}
\usepackage{hyperref}
\usepackage{accents}
\usepackage{bbding}
\usepackage{color,soul}
\usepackage{epstopdf}
\usepackage{changepage}
\usepackage{orcidlink}
\usepackage[normalem]{ulem}
\usepackage{dsfont}

\newcommand{\erf}[1]{Eq.~(\ref{#1})}
\newcommand{\beq}{\begin{equation}}
\newcommand{\eeq}{\end{equation}}
\newcommand{\nn}{\nonumber}

\newcommand{\dg}{^\dagger}

\newcommand{\bra}[1]{\langle{#1}|}
\newcommand{\ket}[1]{|{#1}\rangle}

\newcommand{\op}[2]{\ket{#1}\bra{#2}}

\newcommand{\Tr}{\text{Tr}}

\newcommand{\dd}{{\rm d}}

\newcommand{\ie}{{\em i.e.}}

\newcommand{\past}[1]{\overleftarrow{#1}}
\newcommand{\fut}[1]{\overrightarrow{#1}}
\newcommand{\both}[1]{\overleftrightarrow{#1}}
\newcommand{\fil}{_{\text F}}
\newcommand{\rfil}{_{\text R}}
\newcommand{\sm}{_{\text S}}

\newcommand{\inv}{^{-1}}
\newcommand{\bx}{{\bf x}}

\newcommand{\by}{{\bf y}}

\newcommand{\est}[1]{\check{#1}}

\definecolor{nblue}{rgb}{0.06,0.3,0.73}
\definecolor{nblack}{rgb}{0,0,0}
\definecolor{nred}{rgb}{0.9,0.1,0.1}
\definecolor{nmagenta}{rgb}{0.7,0.0,0.3}
\definecolor{applegreen}{rgb}{0.55, 0.71, 0.0}
\definecolor{nteal}{rgb}{0, 0.5, 0.5}

\newcommand\stPW{\bgroup\markoverwith{\applegreen{\rule[0.5ex]{2pt}{0.4pt}}}\ULon}

\begin{document}

\title{Optimality of the Petz-Fuchs Smoothed State}
\author{Kiarn T. Laverick\orcidlink{0000-0002-3688-1159}}
\affiliation{MajuLab, CNRS-UCA-SU-NUS-NTU International Joint Research Laboratory\\
Centre for Quantum Technologies, National University of Singapore, 117543 Singapore, Singapore}

\date{\today}

\begin{abstract}
Quantum state smoothing is an estimation technique that uses past and future measurement information obtained from a continuous-in-time measurement record to reconstruct the state of an open quantum system. Recently, there has been a new approach to quantum state smoothing based on quantum retrodiciton known as the Petz-Fuchs smoothed state. Here I prove that this Petz-Fuchs smoothed state majorizes all other smoothed states that only depend on the filtered state (the operator conditioned on the past measurement record) and the retrofiltered effect (the operator describing the future measurement record). To do this, I characterize the set of all possible smoothed state satisfying this condition.

\end{abstract}
\pacs{}
\maketitle
\section{Introduction}
Smoothing is a classical estimation technique that uses a continuous-in-time measurement record gathered both prior and posterior to a time $t$ to estimate the state of a classical, Markovian open system \cite{Einicke12,Sarkka13}. As this technique uses information after time $t$ one cannot implement this technique in real time, unlike its counterpart, filtering \cite{Jaz07,BroHwa12}, which only utilizes the measurement record prior to the estimation time (the {\em past} record). In the case of Markovian open quantum systems \cite{LiHallWis18}, one can consider a similar scenario, where one performs a continuous-in-time weak measurement of the environment to gain more information about the state of the system. The concept of filtering (also known as quantum trajectory theory) \cite{Carmichael93,Bel92,Bel99,WisMil10} follows analogously from the classical theory, yielding a conditional quantum state based on the past measurement record. However, directly adapting the classical smoothing theory to the quantum scenario posed problems due to the non-commutative nature of the filtered state $\rho\fil(t)$ and the the retrofiltered effect $\hat{E}\rfil(t)$, an effect operator describing the measurement record posterior to time $t$ (the {\em future} record).

While there were many approaches to utilizing past and future measurement information in quantum systems \cite{Watanabe55,ABL64,Tsa-PRA09,CDJ13,GJM13} (see Ref.~\cite{CGLW21} for a more comprehensive discussion of these techniques), 
recently \cite{LASL25}, there was a new approach to quantum state smoothing proposed.
This approach to smoothing viewed classical smoothing from the perspective of Bayesian retrodiction, whereby the smoothed state is dynamically constructed backwards-in-time from the filtered state at the final time, conditioning on the future record as it evolves. Applying the the quantum analogue of classical Bayesian retrodiction, the Petz recovery map \cite{petz1986,petz1988}, to the filtered quantum state, one obtains the Petz-Fuchs smoothed state,
\beq
\rho\sm^{\rm PF}(t) \propto \sqrt{\rho\fil(t)}\hat{E}\rfil(t)\sqrt{\rho\fil(t)}\,.
\eeq
While this state is a valid smoothed quantum state,
it is unclear in what sense, if any, this state is optimal. As quantum retrodiction is a currently active research field \cite{ManJor19,SS23,PF23,PB23}, with different approach to retrodiction existing, the Petz-Fuchs smoothed state is just one of many possible valid smoothed quantum state, some of which may be a better in terms of, say, purity or entropy. In this paper I will show that the Petz-Fuchs smoothed state majorizes all other valid smoothed quantum states that only depend on the filtered state and retrofiltered effect, and hence is the purest possible smoothed state that one can assign to the system. To do this, I will first characterize the entire set of smoothed quantum states that depend only on the filtered state and retrofiltered effect, from which one can show that the Petz-Fuchs state majorizes all other elements of this set.

The structure of this paper is as follows. First, in Sec.~\ref{Sec:Review}, I will begin by briefly reviewing both classical and quantum state filtering and smoothing, as well as the four criteria that a state must satisfy in order to be consider a valid smoothed quantum state. Next, in Sec.~\ref{Sec:Set}, I will characterize the set of valid smoothed quantum states that only depend on the filtered state and retrofiltered effect. Finally, in Sec.~\ref{Sec:Majorize}, I will show that the Petz-Fuchs smoothed state majorizes the set of valid smoothed quantum states.

\section{Brief Review: Filtering and Smoothing}\label{Sec:Review}
This section will only provide a brief review of the necessary concepts required in this paper. For a more comprehensive review of classical state filtering and smoothing, see Refs.~\cite{Jaz07,Sarkka13}, and for quantum state filtering and smoothing, see Refs.~\cite{WisMil10,LASL25}. Note, here I will only review Petz-Fuchs smoothing; for an in depth review of an alternative formulation, known as Guevara-Wiseman smoothing, see Ref.~\cite{GueWis15,CGLW21}.

\subsection{Classical Systems}
Let us consider an Markovian open classical system described by a vector of parameters $\bx$, which, for simplicity, we will assume can only take discrete values in the set ${\mathbb X}$. The state of this system at time $t$ is defined as the probability distribution of the parameters, $\wp(\bx;t)$. As we are considering a discrete Markovian system, the dynamics are given by
\beq
\wp(\bx;t+\dd t) = \sum_{\bx' \in {\mathbb X}} \varphi(\bx;t+\dd t|\bx';t) \wp(\bx';t)\,,
\eeq
with initial condition $\wp(\bx;t_0)$, where $\varphi(\bx;t+\dd t|\bx';t)$ describing the dynamical transition probabilities for the system. Here, the transition probabilities must satisfy the completeness condition $\sum_{\bx \in {\mathbb X}} \varphi(\bx;t+\dd t|\bx';t) = 1$ to preserve total probability. This state is generally referred to as the unconditioned state, and, in the absence of additional measurement information, is the most accurate description of the system.

To obtain more information about the system one can continuously measure the system obtaining a vector of noisy measurement signals $\by_t$. In the case of filtering, one makes use of the past measure record, defined as $\past{O}_t = \{\by_\tau: \tau \in [t_0, t)\}$ to obtain a more accurate description of the system at time $t + \dd t$. Using Bayes theorem to update our knowledge of the state, the filtered state evolves as
\beq
\tilde\wp\fil(\bx;t+\dd t) = \sum_{\bx' \in {\mathbb X}} {\cal F}_{\by_t}(\bx;t+\dd t|\bx';t) \tilde\wp\fil(\bx';t)\,,
\eeq
where $\wp\fil(\bx;t):=\wp(\bx;t|\past{O}_t)$ and the initial condition is $\wp(\bx;t_0)$. Here the tilde accent $\tilde{\bullet}$ denotes that this state is unnormalized and the conditional dynamical map is ${\cal F}_{y_t}(\bx;t+\dd t|\bx';t) = \varphi_{y_t}(\bx;t+\dd t|\bx';t)\wp(\by_t|\bx;t)$ and the unconditional transition probabilities have been modified to $\varphi_{y_t}(\bx;t+\dd t|\bx';t)$ to account for possible measurement backaction, which can occur even in classical systems. 

Finally, to obtain an even more accurate estimate of the system, one can use the {\em past-future} measurement record, $\both{O} = \{\by_\tau:t\in[t_0,T)\}$ where $T$ is the final time. Taking the Bayesian retrodiction approach, one can obtain the smoothed state via the dynamical equation
\beq
\tilde{\wp}\sm(\bx;t) = \sum_{\bx' \in {\mathbb X}}{\cal R}^{\tilde{\wp}\fil(\bx;t)}_{\by_t}(\bx;t|\bx';t + \dd t)\tilde{\wp}\sm(\bx';t+\dd t)\,,
\eeq
where $\wp\sm(\bx;t):=\wp(\bx;t|\both{O})$ and the final condition is $\tilde\wp\sm(\bx;T):=\tilde\wp\fil(\bx;T)$. The Bayesian retrodictive map \cite{BusSca21} is defined as
\beq
\begin{split}
{\cal R}^{\tilde{\wp}\fil(\bx;t)}_{\by_t}&(\bx;t|\bx';t + \dd t) = \\
&\frac{{\cal F}_{\by_t}(\bx';t+\dd t|\bx;t)\tilde\wp\fil(\bx;t)}{\sum_{\bx''\in{\mathbb X}}{\cal F}_{\by_t}(\bx';t+\dd t|\bx'';t)\tilde\wp\fil(\bx'';t)}\,.
\end{split}
\eeq
Importantly, one should note that the retrodictive map depends on the filtered state. This is because, in retrodiction, one generally needs a reference prior to help invert the forward dynamical map, in this case ${\cal F}_{\by_t}$, in the event that the map is not bijective. Generally, the reference prior is chosen to be one's best guess of the state of the system at time $t$, which in this case is the filtered state. One can write the smoothed state in another form, which is more intuitive from a Bayesian update perspective, by introducing a secondary probability distribution, which I will refer to as the retrofiltered effect for similarity with the quantum terminology, $E\rfil(\bx;t):=\wp(\fut{O}_t|\bx;t)$ where the future record is defined as 
$\fut{O}_t = \{\by_\tau:t\in[t,T)\}$. This retrofiltered effect evolves as
\beq
E\rfil(\bx;t) = \sum_{\bx' \in {\mathbb X}} {\cal F}_{\by_t}(\bx';t+\dd t|\bx;t) E\rfil(\bx';t+\dd t)\,,
\eeq
from a final uninformative state $E\rfil(\bx;T) \propto 1$. With the retrofiltered effect defined, the smoothed state can also be written as 
\beq
\wp\sm(\bx;t) = \frac{E\rfil(\bx;t)\wp\fil(\bx;t)}{\sum_{\bx'\in{\mathbb X}}E\rfil(\bx';t)\wp\fil(\bx';t)}\,.
\eeq

\subsection{Quantum Systems}
Let us consider a Markovian open quantum system, in the strongest sense \cite{LiHallWis18}, where the system is described by a $d$-dimensional density operator $\rho(t)$. The evolution of such a system is given by a Lindblad master equation, however, for simplicity, I will just express this as a completely-positive, trace-preserving map ${\cal E}$. As such, the unconditioned state evolves as
\beq
\rho(t+\dd t) = {\cal E}[\rho(t)]\,,
\eeq
with initial condition $\rho_0$. As with the classical case, this is the most accurate description of the quantum system in the absence of additional information.

One can gain more information about the system by continuously monitoring the environment that the system is coupled with, providing a vector of classical measurement outcomes $\by_t$. Note, the past, future and past-future records are defined identically to the classical case here. Such a measurement of the environment unravels the unconditioned dynamics ${\cal E}$ into a particular Kraus representation according to the measurement operators on the system $\hat{M}_{\by_t}$, $\hat{\cal E}[\bullet] = \sum_{\by_t,i}\hat{M}_{\by_t}\hat{X}_i\bullet\hat{X}_i\dg \hat{M}_{\by_t}\dg$ with the operator $\hat{X}_i$ describing any remaining dissipation that is affecting the system, either due to measurement inefficiency or unmeasured environments. Repeatedly conditioning the quantum state on the measurement outcomes, one obtained the filtered quantum state, which evolves as
\beq
\tilde{\rho}\fil(t+\dd t) = \sum_i \hat{M}_{\by_t}\hat{X}_i\tilde{\rho}\fil(t)\hat{X}_i\dg \hat{M}_{\by_t}\dg := \hat{\cal F}_{\by_t}[\tilde{\rho}\fil(t)]\,,
\eeq
with initial condition $\rho_0$. To describe the future measurement record, one can introduce an effect operator $\hat{E}\rfil(t)$, defined such that $\wp(\fut{O}_t|\rho) = \Tr[\hat{E}\rfil(t)\rho]$. This effect operator evolves via
\beq
\hat{E}\rfil(t) = \hat{\cal F}_{\by_t}\dg[\hat{E\rfil}(t+\dd t)]\,,
\eeq
from a final uninformative state $\hat{E}\rfil(T) = \hat{\mathds 1}$. Note, the retrofiltered effect satisfies the completeness condition $\sum_{\fut{O}_t}\hat{E}\rfil(t) = \hat{\mathds 1}$.

At this point, it should be clear that the filtered state and retrofiltered effect behaves analogously to their classical counterparts. Following this logic, we can apply a quantum retrodictive map analogous to the classical Bayesian map, the Petz recovery map, to obtain a smoothed quantum state,
\beq
\tilde{\rho}^{\rm PF}\sm(t) = \hat{\cal R}_{\by_t}^{\tilde\rho\fil(t)}[\tilde{\rho}^{\rm PF}\sm(t)]\,,
\eeq
where the final condition is taken to be $\tilde\rho\sm^{\rm PF} (T) = \tilde{\rho}\fil(T)$ and the superscript ${\rm PF}$ is to denote that this is the Petz-Fuchs smoothed state. Here, the Petz recovery map is defined as
\beq
\hat{{\cal R}}_{\by_t}^{\gamma}[\bullet] = \sqrt{\gamma}\hat{\cal F}_{\by_t}\dg\left[\sqrt{\hat{\cal F}_{\by_t}\dg[\gamma]\inv} \bullet \sqrt{\hat{\cal F}_{\by_t}\dg[\gamma]\inv}\right]\sqrt{\gamma}\,,
\eeq
where the reference prior $\gamma$ is to be understood in the same way as the classical reference prior. As it turns out, one can solve this dynamical equation analytically \cite{LASL25} in terms of the filtered state and the retrofiltered effect, where one obtains, after renormalization, 
\beq
\rho\sm^{\rm PF}(t) = \frac{\sqrt{\rho\fil(t)}\hat{E}\rfil(t)\sqrt{\rho\fil(t)}}{\Tr[\hat{E}\rfil(t)\rho\fil(t)]}\,.
\eeq
One will note that this form is analogous to the classical smoothed state in terms of the filtered state and retrofiltered effect. 

As mentioned earlier, this is just one of a set of possible smoothed states that one can assign to the system that only depends on the filtered state and retrofiltered effect; another known case is the diagonalized smoothed state (see the Supp. Mat. of Ref.~\cite{LASL25}). As such, it is useful to know what criteria a state must satisfy in order for it to be considered a valid smoothed quantum state. There are 4 criteria, introduced by Wiseman \cite{LCW-QS21}, that a smoothed quantum state should satisfy:
\begin{enumerate}
\item{The theory should give a single smoothed quantum state $\rho\sm$ analogous to the classical state ${\frak S}$, and not a pair of states, for example.}
\item{The smoothed state $\rho\sm$ should reduce to its corresponding filtered state after averaging over all possible
future measurement records $\fut{O}_t$ given a past measurement record, \ie,
\beq
\rho\fil(t) = {\mathbb E}_{\fut{O}_t|\past{O}_t}\{\rho\sm(t)\}\,. \nn
\eeq}
\item{The smoothed quantum state should reduce to its classical counterpart when the initial conditions, final conditions, and dynamics of the system can all be described probabilistically in a fixed basis.}
\item{The smoothed quantum state must be an ${\frak S}$-class quantum state; that is, it must be Hermitian and positive semidefinite with unit trace.}
\end{enumerate}
While each of these criteria are fairly natural, there is some nuance in the third criterion. Specifically, this criteria can be interpreted in different ways depending on the context of the smoothing problem. In our case, it should be interpreted as referring to the filtered state and retrofiltered effect being co-diagonal in a fixed basis. Note, in the case of Guevara-Wiseman smoothing, this condition is strictly stronger, also applying to the true state of the system \cite{LCW-QS21}. With these criteria, one can define, as I will in the next section, the set of valid smoothed states. However, as I desire similarity to the classical smoothed state, being that it only depends on the filtered state and retrofiltered effect, I will limit the set to consider quantum states that can be constructed from these operators.

\section{The Set of Possible Smoothed States}\label{Sec:Set}
Let us first consider the implications of Criteria 1 in the case where we are limiting ourself to the smoothed state only depending on the filtered state and retrofiltered effect. Quite simply, these two conditions together mean any possible smoothed state $\est{\rho}\sm(t)$ can be written as $\est\rho\sm(t) = f\{\rho\fil(t),\hat{E}\rfil(t)\}$, where $f$, at this stage, is some arbitrary function. We will now proceed to constrain the possible functions $f$ based on the remaining criteria to identify the class of all possible smoothed states.

Starting with Criteria 2, $\est\rho\sm(t)$ must satisfy 
\beq
\begin{split}\label{eq:C2_proof}
\rho\fil(t) &= \sum_{\fut{O}_t} \wp(\fut{O}_t|\past{O}_t)\est{\rho}\sm(t)\\
&= \sum_{\fut{O}_t}\tilde f\{\rho\fil(t),\hat{E}\rfil(t)\}\,,
\end{split}
\eeq
where, in the second line, we have used the fact that $\wp(\fut{O}_t|\past{O}_t) = \Tr[\hat{E}\rfil(t)\rho\fil(t)]$ is just a function of $\hat{E}\rfil(t)$ and $\rho\fil(t)$ to define the unnormalized state $\tilde{\est\rho}\sm(t) = \tilde{f}\{\rho\fil(t),\hat{E}\rfil(t)\}$, with $\Tr[\tilde{f}\{\rho\fil(t), \hat{E}\rfil(t)\}] = \Tr[\hat{E}\rfil(t)\rho\fil(t)]$. This provides the first constraint on the function $\tilde{f}$.

For Criteria 3, let us consider a trivial case where the filtered state and retrofiltered effect commute, $\hat{E}\rfil(t) = \hat{\mathds 1}$. By Criteria 3, it must be the case that $\tilde{f}\{\rho\fil(t),\hat{E}\rfil(t)\}$ reproduces the classical smoothed state 
\beq
\rho\sm^{\rm cl} = \sum_x \frac{\wp\fil(x;t)E\rfil(x;t)}{\wp(\fut{O}|\past{O})}\op{\psi_x(t)}{\psi_x(t)},
\eeq
where $\ket{\psi_x(t)}$ is the common basis shared by $\rho\fil(t)$ and $\hat{E}\rfil(t)$ at time $t$. Choosing $E\rfil(x;t) = 1$, consistent with $\hat{E}\rfil(t) = \hat{\mathds 1}$, the function $\tilde{f}$ must also satisfy
\beq\label{eq:function_id}
\tilde{f}\{\rho\fil(t),\hat{\mathds 1}\} = \rho\fil(t)\,,\,\,\,\, \forall \rho\fil(t)\,.
\eeq

Making use of the completion relation for the retrofiltered effect, $\sum_{\fut{O}_t} \hat{E}\rfil(t) = \hat{\mathds 1}$, combining \erf{eq:C2_proof} with \erf{eq:function_id}, we have 
\beq\label{eq:Lin_Constr}
\sum_{\fut{O}_t}\tilde f\{\rho\fil(t),\hat{E}\rfil(t)\} = \tilde{f}\left\{\rho\fil(t), \sum_{\fut{O}_t} \hat{E}\rfil(t)\right\}\,,\,\,\,\, \forall \rho\fil(t)\,.
\eeq
This property is tremendously useful and will drastically limit the space of potential functions ${\tilde f}$. To see this, let's consider the most general expansion possible,
\beq
\begin{split}
\tilde{f}\{\rho\fil(t),\hat{E}\rfil(t)\} = \sum_{i,i',i'',...} &c_{i,i',i'', ...} G_i[\rho\fil(t)]\times \\&G'_{i'}[\hat{E}\rfil(t)]G''_{i''}[\rho\fil(t)]\,...\,,
\end{split}
\eeq
where the set of functions ${\mathbb G} = \{G_i\}$, similarly $\mathbb G'$, {\em etc.}, comprises {\em all} possible functions (discontinuous, basis-dependent, {\em etc}.). Applying \erf{eq:Lin_Constr} to this form heavily limits the possible functions containing $\hat{E}\rfil$ to
\beq
\tilde{f}\{\rho\fil(t),\hat{E}\rfil(t)\} = \sum_{i,j} c_{i,j} G_i[\rho\fil(t)]\hat{E}\rfil(t) H_{j}[\rho\fil(t)]\,,
\eeq
where there are now only two sets of possible functions ${\mathbb G}$ and ${\mathbb H}=\{H_j\}$ remaining. 

Moving onto the final criteria, Criteria 4, our smoothed state must be Hermitian, unit trace and positive semidefinite. Beginning with the Hermiticity, it is easy to see that it must be the case that ${\mathbb G} = {\mathbb H}$. Moreover, if there exists a $G_i\in{\mathbb G}$ such that $G_i\ne G_i\dg$, then $G_i\dg \in{\mathbb G}$. Thus, without loss of generality, we can write the function as 
\beq\label{eq:functional_form}
\tilde{f}\{\rho\fil(t),\hat{E}\rfil(t)\} = \sum_{i,j} c_{i,j} G_i[\rho\fil(t)]\hat{E}\rfil G\dg_{j}[\rho\fil(t)]\,.
\eeq

At this point, we are going to work in the basis that diagonalizes the filtered state, \ie, $\rho\fil(t) \to \sum_{x}\wp\fil(x;t)\op{\psi_x(t)}{\psi_x(t)}$ and $\hat{E}\rfil(t)\to \hat{E}^{(\psi)}\rfil(t)$, where the superscript $(\psi)$ is a reminder that this operator is in the diagonal basis of the filtered state basis $\{\ket{\psi_x}\}$. Substituting this into \erf{eq:functional_form}, we get
\beq
\tilde{f}\{\rho\fil(t),\hat{E}\rfil(t)\} = \sum_{x,z} K_{x,z}(t) E^{(\psi)}_{x,z}(t) \op{\psi_x(t)}{\psi_z(t)}\,,
\eeq
where $K_{x,z}(t) = \sum_{i,j} c_{i,j} G_i[\wp\fil(x;t)] G\dg_{j}[\wp\fil(z;t)]$ and $E^{(\psi)}_{x,z}(t) = \bra{\psi_x(t)}\hat{E}^{(\psi)}\rfil(t)\ket{\psi_z(t)}$. Making use of \erf{eq:function_id}, we see that it must be the case that $K_{x,x} = \wp\fil(x;t)$. We can simplify this computation by defining the Hermitian operator $\hat{K} = \sum_{x,z} K_{x,z} \op{\psi_x(t)}{\psi_z(t)}$, where the smoothed state can be written as
\beq
\tilde{f}\{\rho\fil(t),\hat{E}\rfil(t)\} = \hat{K} \odot \hat{E}\rfil'(t) \,,
\eeq
with $\odot$ being the Hadamard product (or entry-wise matrix product). Here we can see that all of the (unnormalized) functions have the same trace, \ie, $\Tr[\tilde{f}\{\rho\fil(t),\hat{E}\rfil(t)\}] = \sum_x \wp\fil(x;t)E_{x,x}^{(\psi)}(t) = \Tr[\rho\fil(t)E\rfil(t)]$. As the normalized state is defined as $\tilde{f}\{\rho\fil(t),\hat{E}\rfil(t)\}/ \Tr[\rho\fil(t)\hat{E}\rfil(t)]$, this means that $\est{\rho}\sm(t)$ will always be unit trace. For the final part of Criteria 4, that $\tilde{f}\{\rho\fil(t),\hat{E}\rfil(t)\}$ is positive semidefinite for any physical $\hat{E}\rfil(t)$, this forces $\hat{K}\geq 0$. To see this, one only needs to consider the effect operator $\hat{E}\rfil^{(\psi)}(t) = \sum_{x,z}\op{\psi_x(t)}{\psi_z(t)}$.   

At this point, we have identified the entire set of valid smoothed states that that only depend on $\rho\fil(t)$ and $\hat{E}\rfil(t)$ to be 
\beq
\est\rho\sm^{(\psi)}(t) = \frac{\hat{K} \odot \hat{E}\rfil^{(\psi)}(t)}{\Tr[\rho\fil(t)\hat{E}\rfil(t)]}\,,
\eeq
where $\hat{K}\geq 0$ and $K_{x,x}(t) = \wp\fil(x;t)$, where I remind the reader that the prime denotes that an operator is in represented in the diagonal basis of the filtered state at time $t$.

\section{Optimality of the Petz-Fuchs Smoothed State}\label{Sec:Majorize}
Now that we have identified the set of valid smoothed quantum state that satisfy the smoothing criteria, we can show that the Petz-Fuchs smoothed state majorizes this set. Before I prove this statement, I will remind the reader what it means for a state $\rho$ to majorize another state $\sigma$, denoted as $\rho\succ\sigma$. Let us assume that $\rho$ and $\sigma$ are $d$-dimensional quantum states. $\rho$ is said to {\em majorize} \cite{Watrous18} $\sigma$ if $\sum_{i = 0}^m \lambda_i(\rho) >\sum_{i = 0}^{m}\lambda_i(\sigma)$ for all $m\leq d$, where $\lambda_i(A)$ denotes the $i$th eigenvalues of $A$ in decreasing order, \ie, $\lambda_i(A)\geq\lambda_{i+1}(A)$. Importantly, it should be noted that the following two statements are equivalent \cite{Watrous18}:
\begin{itemize}
    \item[(i)] $\rho \succ \sigma$.
    \item[(ii)] There exists a unital channel $\Phi$ the turns $\rho$ into $\sigma$, \ie, $\sigma = \Phi[\rho]$.
\end{itemize}

With the preliminaries out of the way, I will now show that the set of valid smoothed states can be expressed as $\est{\rho}\sm(t) = \Phi[\rho^{\rm PF}\sm(t)]$ where $\Phi$ is a unital channel. First note that one obtains the Petz-Fuchs smoothed state in the diagonal basis of the filtered state when $K_{x,z} = \sqrt{\wp\fil(x;t)\wp\fil(z;t)}$. Let us define the correlation matrix $C_{x,z} = K_{x,z}/\sqrt{\wp\fil(x;t)\wp\fil(z;t)}$, where a correlation matrix is defined as a symmetric positive semidefinite  with unit elements on the diagonal, \ie, $C\geq 0$ and $C_{x,x} = 1$. It is easy to show that the correlation matrix $C_{x,z} = K_{x,z}/\sqrt{\wp\fil(x;t)\wp\fil(z;t)}$ satisfies both of these conditions. With this correlation matrix defined we can now express any valid smoothed state as $\est\rho\sm^{(\psi)}(t) = C\odot\rho\sm^{\rm PF (\psi)}(t)$. Since the Hadamard product between a quantum state and a correlation matrix is equivalent to a unital channel $\Phi[\bullet]$ applied to $\rho\sm^{\rm PF(\psi)}(t)$ \cite{Watrous18}, we have that $\est\rho\sm^{(\psi)}(t) = \Phi[\rho\sm^{\rm PF(\psi)}(t)]$. Since unitary transformations of a unital channel are result in another unital channel, \ie, $U\Phi[V\bullet V\dg]U\dg = \Phi'[\bullet]$ for any unitaries $U$ and $V$, we can move out of the diagonal basis of the filtered state to any basis. This gives that any valid smoothed state that is a function of $\rho\fil(t)$ and $\hat{E}\rfil(t)$ is given by
\beq
\est\rho\sm(t) = \Phi[\rho^{\rm PF}\sm(t)]\,,
\eeq
where $\Phi$ is a unital channel, which is equivalent to $\rho\sm^{\rm PF}(t) \succ \est\rho\sm(t)$.

An immediate consequence of the fact that the Petz-Fuch smoothed state majorizes all valid smoothed states means that any Schur concave function of $\rho\sm^{\rm PF}$ will be minimized, like the von Neumann, R\'enyi entropy. Similarly, any Schur convex will be maximized, like the purity. This means that, the Petz-Fuch state will be the purest smoothed assignment that one can attribute to an open Markovian quantum system given a past and future measurement record. Another way to understand this is that the Petz-Fuch smoothed state uses maximally utilises the information contained in the past-future measurement record to reconstruct the purest description of the system. It should be noted, however, that on a individual trajectory level, this does not mean that the Petz-Fuchs smoothed state (or any valid smoothed state for that matter) will be purer than the filtered state (See Ref.~\cite{LASL25} for an example). Nevertheless, on average, the smoothed state always performs better than the filtered state \cite{LASL25}. 

\section{Conclusion}
In this paper, I have characterized the complete set of possible valid smoothed states that only depend on the filtered state and retrofiltered effect, similar to the classical case. Subsequently, I have shown that the Pet-Fuchs smoothed state majorizes all other smoothed states in this set, showing that it is the purest possible smoothed reconstruction of the system state. As this is the purest assignment possible to an open quantum system given past and future measurement information, it would be interesting to see whether this knowledge of the system can help in understanding particular physical phenomenon better. As an example, could one learn better the state of a light field scattered of a atom coupled to a bosonic bath.


\acknowledgements
I would like to thank Mingxuan Liu, Samyak P. Prasad, Kian Hwee Lim, Nathan Shettell and Howard M. Wiseman for discussions on the initial idea of this paper. This project is supported by the National Research Foundation, Singapore through the National Quantum Office, hosted in A*STAR, under its Centre for Quantum Technologies Funding Initiative (S24Q2d0009), and the Plan France 2030 through the projects NISQ2LSQ (Grant ANR-22-PETQ-0006), OQuLus (Grant ANR-22-PETQ-0013), and OECQ through BPI France. 


\begin{thebibliography}{26}%
\makeatletter
\providecommand \@ifxundefined [1]{%
 \@ifx{#1\undefined}
}%
\providecommand \@ifnum [1]{%
 \ifnum #1\expandafter \@firstoftwo
 \else \expandafter \@secondoftwo
 \fi
}%
\providecommand \@ifx [1]{%
 \ifx #1\expandafter \@firstoftwo
 \else \expandafter \@secondoftwo
 \fi
}%
\providecommand \natexlab [1]{#1}%
\providecommand \enquote  [1]{``#1''}%
\providecommand \bibnamefont  [1]{#1}%
\providecommand \bibfnamefont [1]{#1}%
\providecommand \citenamefont [1]{#1}%
\providecommand \href@noop [0]{\@secondoftwo}%
\providecommand \href [0]{\begingroup \@sanitize@url \@href}%
\providecommand \@href[1]{\@@startlink{#1}\@@href}%
\providecommand \@@href[1]{\endgroup#1\@@endlink}%
\providecommand \@sanitize@url [0]{\catcode `\\12\catcode `\$12\catcode
  `\&12\catcode `\#12\catcode `\^12\catcode `\_12\catcode `\%12\relax}%
\providecommand \@@startlink[1]{}%
\providecommand \@@endlink[0]{}%
\providecommand \url  [0]{\begingroup\@sanitize@url \@url }%
\providecommand \@url [1]{\endgroup\@href {#1}{\urlprefix }}%
\providecommand \urlprefix  [0]{URL }%
\providecommand \Eprint [0]{\href }%
\providecommand \doibase [0]{http://dx.doi.org/}%
\providecommand \selectlanguage [0]{\@gobble}%
\providecommand \bibinfo  [0]{\@secondoftwo}%
\providecommand \bibfield  [0]{\@secondoftwo}%
\providecommand \translation [1]{[#1]}%
\providecommand \BibitemOpen [0]{}%
\providecommand \bibitemStop [0]{}%
\providecommand \bibitemNoStop [0]{.\EOS\space}%
\providecommand \EOS [0]{\spacefactor3000\relax}%
\providecommand \BibitemShut  [1]{\csname bibitem#1\endcsname}%
\let\auto@bib@innerbib\@empty
\bibitem [{\citenamefont {Einicke}(2012)}]{Einicke12}%
  \BibitemOpen
  \bibfield  {author} {\bibinfo {author} {\bibfnamefont {G.~A.}\ \bibnamefont
  {Einicke}},\ }\href@noop {} {\emph {\bibinfo {title} {Smoothing, filtering
  and prediction: Estimating the past, present and future}}}\ (\bibinfo
  {publisher} {InTech},\ \bibinfo {address} {Rijeka Croatia},\ \bibinfo {year}
  {2012})\BibitemShut {NoStop}%
\bibitem [{\citenamefont {S{\"a}rkk{\"a}}(2013)}]{Sarkka13}%
  \BibitemOpen
  \bibfield  {author} {\bibinfo {author} {\bibfnamefont {S.}~\bibnamefont
  {S{\"a}rkk{\"a}}},\ }\href@noop {} {\emph {\bibinfo {title} {Bayesian
  filtering and smoothing}}},\ Vol.~\bibinfo {volume} {3}\ (\bibinfo
  {publisher} {Cambridge University Press},\ \bibinfo {address} {Cambridge,
  England},\ \bibinfo {year} {2013})\BibitemShut {NoStop}%
\bibitem [{\citenamefont {Jazwinski}(2007)}]{Jaz07}%
  \BibitemOpen
  \bibfield  {author} {\bibinfo {author} {\bibfnamefont {A.~H.}\ \bibnamefont
  {Jazwinski}},\ }\href@noop {} {\emph {\bibinfo {title} {Stochastic processes
  and filtering theory}}}\ (\bibinfo  {publisher} {Dover Publications},\
  \bibinfo {address} {New York},\ \bibinfo {year} {2007})\BibitemShut {NoStop}%
\bibitem [{\citenamefont {Brown}\ and\ \citenamefont {Hwang}(2012)}]{BroHwa12}%
  \BibitemOpen
  \bibfield  {author} {\bibinfo {author} {\bibfnamefont {R.~G.}\ \bibnamefont
  {Brown}}\ and\ \bibinfo {author} {\bibfnamefont {P.~Y.~C.}\ \bibnamefont
  {Hwang}},\ }\href@noop {} {\emph {\bibinfo {title} {{Introduction to random
  signals and applied Kalman filtering}}}},\ \bibinfo {edition} {4th}\ ed.\
  (\bibinfo  {publisher} {Wiley},\ \bibinfo {address} {New York},\ \bibinfo
  {year} {2012})\BibitemShut {NoStop}%
\bibitem [{\citenamefont {Li}\ \emph {et~al.}(2018)\citenamefont {Li},
  \citenamefont {Hall},\ and\ \citenamefont {Wiseman}}]{LiHallWis18}%
  \BibitemOpen
  \bibfield  {author} {\bibinfo {author} {\bibfnamefont {L.}~\bibnamefont
  {Li}}, \bibinfo {author} {\bibfnamefont {M.~J.~W.}\ \bibnamefont {Hall}}, \
  and\ \bibinfo {author} {\bibfnamefont {H.~M.}\ \bibnamefont {Wiseman}},\
  }\bibfield  {title} {\enquote {\bibinfo {title} {{Concepts of quantum
  non-Markovianity: A hierarchy}},}\ }\href {\doibase
  10.1016/j.physrep.2018.07.001} {\bibfield  {journal} {\bibinfo  {journal}
  {Phys. Rep.}\ }\textbf {\bibinfo {volume} {759}},\ \bibinfo {pages} {1--51}
  (\bibinfo {year} {2018})}\BibitemShut {NoStop}%
\bibitem [{\citenamefont {Carmichael}(1993)}]{Carmichael93}%
  \BibitemOpen
  \bibfield  {author} {\bibinfo {author} {\bibfnamefont {H.~J.}\ \bibnamefont
  {Carmichael}},\ }\href@noop {} {\emph {\bibinfo {title} {{An open systems
  approach to quantum optics}}}}\ (\bibinfo  {publisher} {Springer},\ \bibinfo
  {address} {Berlin},\ \bibinfo {year} {1993})\BibitemShut {NoStop}%
\bibitem [{\citenamefont {Belavkin}(1992)}]{Bel92}%
  \BibitemOpen
  \bibfield  {author} {\bibinfo {author} {\bibfnamefont {V.~P.}\ \bibnamefont
  {Belavkin}},\ }\bibfield  {title} {\enquote {\bibinfo {title} {{Quantum
  continual measurements and a posteriori collapse on CCR}},}\ }\href {\doibase
  10.1007/BF02097018} {\bibfield  {journal} {\bibinfo  {journal} {Commun. Math.
  Phys.}\ }\textbf {\bibinfo {volume} {146}},\ \bibinfo {pages} {611--635}
  (\bibinfo {year} {1992})}\BibitemShut {NoStop}%
\bibitem [{\citenamefont {Belavkin}(1999)}]{Bel99}%
  \BibitemOpen
  \bibfield  {author} {\bibinfo {author} {\bibfnamefont {V.~P.}\ \bibnamefont
  {Belavkin}},\ }\bibfield  {title} {\enquote {\bibinfo {title} {{Measurement,
  filtering and control in quantum open dynamical systems}},}\ }\href {\doibase
  doi:10.1016/S0034-4877(00)86386-7} {\bibfield  {journal} {\bibinfo  {journal}
  {Rep. Math. Phys.}\ }\textbf {\bibinfo {volume} {43}},\ \bibinfo {pages}
  {A405} (\bibinfo {year} {1999})}\BibitemShut {NoStop}%
\bibitem [{\citenamefont {Wiseman}\ and\ \citenamefont
  {Milburn}(2010)}]{WisMil10}%
  \BibitemOpen
  \bibfield  {author} {\bibinfo {author} {\bibfnamefont {H.~M.}\ \bibnamefont
  {Wiseman}}\ and\ \bibinfo {author} {\bibfnamefont {G.~J.}\ \bibnamefont
  {Milburn}},\ }\href@noop {} {\emph {\bibinfo {title} {{Quantum measurement
  and control}}}}\ (\bibinfo  {publisher} {Cambridge University Press},\
  \bibinfo {address} {Cambridge, England},\ \bibinfo {year} {2010})\BibitemShut
  {NoStop}%
\bibitem [{\citenamefont {Watanabe}(1955)}]{Watanabe55}%
  \BibitemOpen
  \bibfield  {author} {\bibinfo {author} {\bibfnamefont {S.}~\bibnamefont
  {Watanabe}},\ }\bibfield  {title} {\enquote {\bibinfo {title} {{Symmetry of
  physical laws. Part III. Prediction and retrodiction}},}\ }\href {\doibase
  10.1103/RevModPhys.27.179} {\bibfield  {journal} {\bibinfo  {journal} {Rev.
  Mod. Phys.}\ }\textbf {\bibinfo {volume} {27}},\ \bibinfo {pages} {179}
  (\bibinfo {year} {1955})}\BibitemShut {NoStop}%
\bibitem [{\citenamefont {Aharonov}\ \emph {et~al.}(1964)\citenamefont
  {Aharonov}, \citenamefont {Bergmann},\ and\ \citenamefont
  {Lebowitz}}]{ABL64}%
  \BibitemOpen
  \bibfield  {author} {\bibinfo {author} {\bibfnamefont {Y.}~\bibnamefont
  {Aharonov}}, \bibinfo {author} {\bibfnamefont {P.~G.}\ \bibnamefont
  {Bergmann}}, \ and\ \bibinfo {author} {\bibfnamefont {J.~L.}\ \bibnamefont
  {Lebowitz}},\ }\bibfield  {title} {\enquote {\bibinfo {title} {{Time symmetry
  in the Quantum process of measurement}},}\ }\href {\doibase
  10.1103/PhysRev.134.B1410} {\bibfield  {journal} {\bibinfo  {journal} {Phys.
  Rev.}\ }\textbf {\bibinfo {volume} {134}},\ \bibinfo {pages} {B1410--B1416}
  (\bibinfo {year} {1964})}\BibitemShut {NoStop}%
\bibitem [{\citenamefont {Tsang}(2009)}]{Tsa-PRA09}%
  \BibitemOpen
  \bibfield  {author} {\bibinfo {author} {\bibfnamefont {M.}~\bibnamefont
  {Tsang}},\ }\bibfield  {title} {\enquote {\bibinfo {title} {Optimal waveform
  estimation for classical and quantum systems via time-symmetric smoothing},}\
  }\href {\doibase 10.1103/PhysRevA.80.033840} {\bibfield  {journal} {\bibinfo
  {journal} {Phys. Rev. A}\ }\textbf {\bibinfo {volume} {80}},\ \bibinfo
  {pages} {033840} (\bibinfo {year} {2009})}\BibitemShut {NoStop}%
\bibitem [{\citenamefont {Chantasri}\ \emph {et~al.}(2013)\citenamefont
  {Chantasri}, \citenamefont {Dressel},\ and\ \citenamefont {Jordan}}]{CDJ13}%
  \BibitemOpen
  \bibfield  {author} {\bibinfo {author} {\bibfnamefont {A.}~\bibnamefont
  {Chantasri}}, \bibinfo {author} {\bibfnamefont {J.}~\bibnamefont {Dressel}},
  \ and\ \bibinfo {author} {\bibfnamefont {A.~N.}\ \bibnamefont {Jordan}},\
  }\bibfield  {title} {\enquote {\bibinfo {title} {Action principle for
  continuous quantum measurement},}\ }\href {\doibase
  10.1103/PhysRevA.88.042110} {\bibfield  {journal} {\bibinfo  {journal} {Phys.
  Rev. A}\ }\textbf {\bibinfo {volume} {88}},\ \bibinfo {pages} {042110}
  (\bibinfo {year} {2013})}\BibitemShut {NoStop}%
\bibitem [{\citenamefont {Gammelmark}\ \emph {et~al.}(2013)\citenamefont
  {Gammelmark}, \citenamefont {Julsgaard},\ and\ \citenamefont
  {M\o{}lmer}}]{GJM13}%
  \BibitemOpen
  \bibfield  {author} {\bibinfo {author} {\bibfnamefont {S.}~\bibnamefont
  {Gammelmark}}, \bibinfo {author} {\bibfnamefont {B.}~\bibnamefont
  {Julsgaard}}, \ and\ \bibinfo {author} {\bibfnamefont {K.}~\bibnamefont
  {M\o{}lmer}},\ }\bibfield  {title} {\enquote {\bibinfo {title} {{Past quantum
  states of a monitored system}},}\ }\href {\doibase
  10.1103/PhysRevLett.111.160401} {\bibfield  {journal} {\bibinfo  {journal}
  {Phys. Rev. Lett.}\ }\textbf {\bibinfo {volume} {{\bf 111}}},\ \bibinfo
  {pages} {160401} (\bibinfo {year} {2013})}\BibitemShut {NoStop}%
\bibitem [{\citenamefont {Chantasri}\ \emph {et~al.}(2021)\citenamefont
  {Chantasri}, \citenamefont {Guevara}, \citenamefont {Laverick},\ and\
  \citenamefont {Wiseman}}]{CGLW21}%
  \BibitemOpen
  \bibfield  {author} {\bibinfo {author} {\bibfnamefont {A.}~\bibnamefont
  {Chantasri}}, \bibinfo {author} {\bibfnamefont {I.}~\bibnamefont {Guevara}},
  \bibinfo {author} {\bibfnamefont {K.~T.}\ \bibnamefont {Laverick}}, \ and\
  \bibinfo {author} {\bibfnamefont {H.~M.}\ \bibnamefont {Wiseman}},\
  }\bibfield  {title} {\enquote {\bibinfo {title} {Unifying theory of quantum
  state estimation using past and future information},}\ }\href {\doibase
  10.1016/j.physrep.2021.07.003} {\bibfield  {journal} {\bibinfo  {journal}
  {Phys. Rep.}\ }\textbf {\bibinfo {volume} {930}},\ \bibinfo {pages} {1--40}
  (\bibinfo {year} {2021})}\BibitemShut {NoStop}%
\bibitem [{\citenamefont {Liu}\ \emph {et~al.}(2025)\citenamefont {Liu},
  \citenamefont {Scarani}, \citenamefont {Auff{\`e}ves},\ and\ \citenamefont
  {Laverick}}]{LASL25}%
  \BibitemOpen
  \bibfield  {author} {\bibinfo {author} {\bibfnamefont {M.}~\bibnamefont
  {Liu}}, \bibinfo {author} {\bibfnamefont {V.}~\bibnamefont {Scarani}},
  \bibinfo {author} {\bibfnamefont {A.}~\bibnamefont {Auff{\`e}ves}}, \ and\
  \bibinfo {author} {\bibfnamefont {K.~T.}\ \bibnamefont {Laverick}},\
  }\bibfield  {title} {\enquote {\bibinfo {title} {Retrodictive approach to
  quantum state smoothing},}\ }\href {\doibase 10.1103/8pc3-7pg5} {\bibfield
  {journal} {\bibinfo  {journal} {Phys. Rev. A}\ }\textbf {\bibinfo {volume}
  {112}},\ \bibinfo {pages} {L030203} (\bibinfo {year} {2025})}\BibitemShut
  {NoStop}%
\bibitem [{\citenamefont {Petz}(1986)}]{petz1986}%
  \BibitemOpen
  \bibfield  {author} {\bibinfo {author} {\bibfnamefont {D.}~\bibnamefont
  {Petz}},\ }\bibfield  {title} {\enquote {\bibinfo {title} {Sufficient
  subalgebras and the relative entropy of states of a von neumann algebra},}\
  }\href {\doibase 10.1007/BF01212345} {\bibfield  {journal} {\bibinfo
  {journal} {Commun. Math. Phys.}\ }\textbf {\bibinfo {volume} {105}},\
  \bibinfo {pages} {123--131} (\bibinfo {year} {1986})}\BibitemShut {NoStop}%
\bibitem [{\citenamefont {Petz}(1988)}]{petz1988}%
  \BibitemOpen
  \bibfield  {author} {\bibinfo {author} {\bibfnamefont {D.}~\bibnamefont
  {Petz}},\ }\bibfield  {title} {\enquote {\bibinfo {title} {Sufficiency of
  channels over von {N}eumann algebra},}\ }\href {\doibase
  10.1093/qmath/39.1.97} {\bibfield  {journal} {\bibinfo  {journal} {Q. J.
  Math.}\ }\textbf {\bibinfo {volume} {39}},\ \bibinfo {pages} {97--108}
  (\bibinfo {year} {1988})}\BibitemShut {NoStop}%
\bibitem [{\citenamefont {Manikandan}\ and\ \citenamefont
  {Jordan}(2019)}]{ManJor19}%
  \BibitemOpen
  \bibfield  {author} {\bibinfo {author} {\bibfnamefont {S.~K.}\ \bibnamefont
  {Manikandan}}\ and\ \bibinfo {author} {\bibfnamefont {A.~N.}\ \bibnamefont
  {Jordan}},\ }\bibfield  {title} {\enquote {\bibinfo {title} {Time reversal
  symmetry of generalized quantum measurements with past and future boundary
  conditions},}\ }\href {\doibase 10.1007/s40509-019-00182-w} {\bibfield
  {journal} {\bibinfo  {journal} {Quantum Stud.: Math. Found.}\ }\textbf
  {\bibinfo {volume} {6}},\ \bibinfo {pages} {241--268} (\bibinfo {year}
  {2019})}\BibitemShut {NoStop}%
\bibitem [{\citenamefont {Surace}\ and\ \citenamefont {Scandi}(2023)}]{SS23}%
  \BibitemOpen
  \bibfield  {author} {\bibinfo {author} {\bibfnamefont {J.}~\bibnamefont
  {Surace}}\ and\ \bibinfo {author} {\bibfnamefont {M.}~\bibnamefont
  {Scandi}},\ }\bibfield  {title} {\enquote {\bibinfo {title} {State retrieval
  beyond {B}ayes' retrodiction},}\ }\href {\doibase 10.22331/q-2023-04-27-990}
  {\bibfield  {journal} {\bibinfo  {journal} {{Quantum}}\ }\textbf {\bibinfo
  {volume} {7}},\ \bibinfo {pages} {990} (\bibinfo {year} {2023})}\BibitemShut
  {NoStop}%
\bibitem [{\citenamefont {Parzygnat}\ and\ \citenamefont
  {Fullwood}(2023)}]{PF23}%
  \BibitemOpen
  \bibfield  {author} {\bibinfo {author} {\bibfnamefont {A.~J.}\ \bibnamefont
  {Parzygnat}}\ and\ \bibinfo {author} {\bibfnamefont {J.}~\bibnamefont
  {Fullwood}},\ }\bibfield  {title} {\enquote {\bibinfo {title} {From
  time-reversal symmetry to quantum bayes' rules},}\ }\href {\doibase
  10.1103/PRXQuantum.4.020334} {\bibfield  {journal} {\bibinfo  {journal} {PRX
  Quantum}\ }\textbf {\bibinfo {volume} {4}},\ \bibinfo {pages} {020334}
  (\bibinfo {year} {2023})}\BibitemShut {NoStop}%
\bibitem [{\citenamefont {Parzygnat}\ and\ \citenamefont
  {Buscemi}(2023)}]{PB23}%
  \BibitemOpen
  \bibfield  {author} {\bibinfo {author} {\bibfnamefont {A.~J.}\ \bibnamefont
  {Parzygnat}}\ and\ \bibinfo {author} {\bibfnamefont {F.}~\bibnamefont
  {Buscemi}},\ }\bibfield  {title} {\enquote {\bibinfo {title} {Axioms for
  retrodiction: achieving time-reversal symmetry with a prior},}\ }\href
  {\doibase 10.22331/q-2023-05-23-1013} {\bibfield  {journal} {\bibinfo
  {journal} {{Quantum}}\ }\textbf {\bibinfo {volume} {7}},\ \bibinfo {pages}
  {1013} (\bibinfo {year} {2023})}\BibitemShut {NoStop}%
\bibitem [{\citenamefont {Guevara}\ and\ \citenamefont
  {Wiseman}(2015)}]{GueWis15}%
  \BibitemOpen
  \bibfield  {author} {\bibinfo {author} {\bibfnamefont {I.}~\bibnamefont
  {Guevara}}\ and\ \bibinfo {author} {\bibfnamefont {H.~M.}\ \bibnamefont
  {Wiseman}},\ }\bibfield  {title} {\enquote {\bibinfo {title} {Quantum state
  smoothing},}\ }\href {\doibase 10.1103/PhysRevLett.115.180407} {\bibfield
  {journal} {\bibinfo  {journal} {Phys. Rev. Lett.}\ }\textbf {\bibinfo
  {volume} {115}},\ \bibinfo {pages} {180407} (\bibinfo {year}
  {2015})}\BibitemShut {NoStop}%
\bibitem [{\citenamefont {Buscemi}\ and\ \citenamefont
  {Scarani}(2021)}]{BusSca21}%
  \BibitemOpen
  \bibfield  {author} {\bibinfo {author} {\bibfnamefont {F.}~\bibnamefont
  {Buscemi}}\ and\ \bibinfo {author} {\bibfnamefont {V.}~\bibnamefont
  {Scarani}},\ }\bibfield  {title} {\enquote {\bibinfo {title} {Fluctuation
  theorems from bayesian retrodiction},}\ }\href {\doibase
  10.1103/PhysRevE.103.052111} {\bibfield  {journal} {\bibinfo  {journal}
  {Phys. Rev. E}\ }\textbf {\bibinfo {volume} {103}},\ \bibinfo {pages}
  {052111} (\bibinfo {year} {2021})}\BibitemShut {NoStop}%
\bibitem [{\citenamefont {Laverick}\ \emph {et~al.}(2021)\citenamefont
  {Laverick}, \citenamefont {Chantasri},\ and\ \citenamefont
  {Wiseman}}]{LCW-QS21}%
  \BibitemOpen
  \bibfield  {author} {\bibinfo {author} {\bibfnamefont {K.~T.}\ \bibnamefont
  {Laverick}}, \bibinfo {author} {\bibfnamefont {A.}~\bibnamefont {Chantasri}},
  \ and\ \bibinfo {author} {\bibfnamefont {H.~M.}\ \bibnamefont {Wiseman}},\
  }\bibfield  {title} {\enquote {\bibinfo {title} {{General criteria for
  quantum state smoothing with necessary and sufficient criteria for linear
  Gaussian quantum systems}},}\ }\href {\doibase 10.1007/s40509-020-00225-7}
  {\bibfield  {journal} {\bibinfo  {journal} {Quantum Stud.: Math. Found.}\
  }\textbf {\bibinfo {volume} {8}},\ \bibinfo {pages} {37--50} (\bibinfo {year}
  {2021})}\BibitemShut {NoStop}%
\bibitem [{\citenamefont {Watrous}\ \emph {et~al.}(2018)\citenamefont {Watrous}
  \emph {et~al.}}]{Watrous18}%
  \BibitemOpen
  \bibfield  {author} {\bibinfo {author} {\bibfnamefont {J.}~\bibnamefont
  {Watrous}} \emph {et~al.},\ }\href@noop {} {\emph {\bibinfo {title} {The
  theory of quantum information}}},\ Vol.~\bibinfo {volume} {1}\ (\bibinfo
  {publisher} {Cambridge university press Cambridge},\ \bibinfo {year}
  {2018})\BibitemShut {NoStop}%
\end{thebibliography}
%

\end{document}